\documentstyle[aps,prl]{revtex}
\begin{document}
\newcommand{\dv}{\Delta {\rm v}}
\newcommand{\dbv}{\Delta {\bf v}}
\newcommand{\dbd}{\Delta {\bf d}}
\newcommand{\vE}{{\bf E}}
\newcommand{\vB}{{\bf B}}
\newcommand{\vd}{{\bf d}}
\newcommand{\vR}{{\bf R}}
\newcommand{\vcr}{{\bf r}}
\newcommand{\vv}{{\bf v}}
\newcommand{\pal}{\alpha_{\perp}}
\newcommand{\al}{\alpha}
\newcommand{\be}{\begin{equation}}
\newcommand{\ee}{\end{equation}}
\newcommand{\bea}{\begin{eqnarray}}
\newcommand{\eea}{\end{eqnarray}}
\title
{ Hall-Effect for Neutral Atoms.}

\author
{A.P. Kasantsev $^a$, A.M. Dykhne $^b$ and V.L. Pokrovsky $^a$$^c$.}
\address{$^a$ Landau Institute for Theoretical Physics, Chernogolovka 142432,
Russia.\\
$^b$ Filial of the Kurchatov Institute of the Atomic Energy, Troitsk       ,
Russia.\\
$^c$ Dept. of Physics, Texas A\&M University, College Station, TX 77843-4242,
USA}
\date{\today}
\maketitle
\begin{abstract}
It is shown that polarizable neutral systems can drift in crossed magnetic and
electric field. The drift velocity is perpendicular to both fields, but, 
contrary to
the drift velocity of a charged particle, it exists only if fields vary in space
or in time. We develop an adiabatic theory of this phenomenon and analyze 
conditions of its experimental observation.
\end{abstract}

This work has been designed
and partly performed several years ago. It has been interrupred by
the untimely death of Alexander Petrovich Kasantsev whose contribution 
to this article
was decisive. By completing this work we pay our tribute to the memory of 
our dear friend, a profound scientist, a sincere and kind man.\\

{\it Polarizable system in crossed fields.}
To our knowledge nobody considered the motion of a neutral particles in crossed
electric and magnetic fields. The purpose of this work is to point out that
neutral systems perform a drift in crossed fields, if these systems are 
polarizable.
Consider a neutral system of point particles in external fields.
Equations of motions for such a system are:
\be
m_i\ddot{\vcr }_i = e_i\vE(\vcr_i)+\frac{e_i}{c}\dot{\vcr }_i\times\vB(\vcr_i) -
\sum_{j\neq i}\nabla V_i(\vcr_{ij})
\ee
Notations are obvious. Assuming the electric and magnetic fields to be classical,
equation (1) is valid as an operator equation for the quantized motion of 
the neutral system.
\par
Let the system size to be much smaller than the characteristic length $L$ of the
fields variation. Summing both sides of eqn (1) over $i$ and expanding
$\vE(\vcr_i)$ and $\vB(\vcr_i)$ over small deviations $\vcr_i - \vR$ 
from the center-of-mass position $\vR$, we obtain in the leading approximation:
\be
M\ddot{\vR}=(\vd\nabla)\vE(\vR)+ \frac{\dot{\vd}\times\vB(\vR)}{c}
\ee
Where $\vd=\sum {e_i\vcr_i}$ is the operator of dipolar momentum and
$M=\sum{m_i}$ is the total mass.
\par
Further we accept the adiabatic approximation, i.e., 
we assume that the quantum state of our system adiabatically
follows instant values of fields. It means that the center-of mass motion is
slow enough to provide the characteristic time of motion to be
much larger than the inverse frequency of the system. Quantitative implications
of the adiabaticity requirement will be considered later. We additionally assume
$B\gg E$. As a consequence of the adiabaticity the dipolar momentum is 
a linear function of the field $\vE (\vR )$ at the center-of-mass location:
\be
\vd = \hat{\al}\vE (\vR )
\ee
The polarizability tensor $\hat{\al}$ may depend on magnetic field $\vB (\vR )$.
Thus, a closed differential equation for the center-of-mass radius-vector $\vR$
coincides formally with equation (2), in which $\vd$ is defined by eqn (3) and
$\hat{\al}(\vB )$ is a given tensorial function of $\vB$.
\par
To get a tangible result
we assume the velocity of the center-of-mass $\vv_0 =\dot{\vR_0}$ to be large,
so that the change of velocity $\dbv$ caused by the fields $\vE$ and $\vB$ is 
small compared to $\vv_0$. Then we find for $\dbv$:
\be
\dbv = \frac{1}{M}\int\left[ (\vd\nabla )\vE (\vR (t)) +
\frac{\dot{\vd} \times \vB (\vR )}{c}\right] dt
\ee
where $\vR = \vR (t) =\vR_0 + \vv_{0}t$. Below several possible configurations 
are considered.\\
{\it Constant Fields.}
 First we consider static fields, varying in space.
Let us focus on the second term in the r.-h.s. of eqn (4).
A simplest possible geometry occurs when a condenser is located
inside a solenoid in such a way that the electric field is
perpendicular to the magnetic field. The beam of neutral atoms is supposed
to propagate
along the axis of the solenoid, parallel to the magnetic field and
perpendicular  to the electric field. The magnetic
field $\vB$ is assumed to be a constant. Then:
\be
\dbv = \frac{\dbd \times \vB}{Mc}
\ee
where $\dbd$ is the total change of the dipolar momentum. If the electric 
field $\vE$
is the same at the beginning and at the end of trajectory, the total 
effect is zero.
It is the case for an arbitrary electric field distribution, if it is 
fully located
in the volume occupied by the magnetic  field. Therefore, a more complex 
configuration
is necessary. In order to obtain a finite drift velocity, one can shift
the condenser in such a way that one its end, where the atomic beam enters the
condenser, is located beyond the volume occupied by magnetic field,
whereas the second end is located inside this volume. Then:
\be
\dbv = -\frac{\pal (B)(\vE \times \vB )}{Mc}.
\label{end}
\ee
A  modification of this idea  is to excite the Rydberg
state by a laser beam through a hole in the condenser.
The effect in such a configuration differs by the sign from that
given by Eq. (\ref{end}). In the framework of this theory it does not depend
on the position of a hole along the atomic beam path.
\par
Another simple configuration is delivered by a variation of the magnetic field
in space. Let us direct $x$-axis along the atomic beam and the magnetic
field, $y$-axis along the electric field. We assume that $\vE (x)$ is a constant
in the condenser, whereas $B_{x}(x)$ changes from a value $B_1$ at a point where
the beam enters the condenser till a value $B_2$ at a point where the beam
leaves the condenser \footnote{Since $\nabla \bf B = 0$, a purely longitudinal
variation of $\bf B$ is impossible and transverse components are inavoidable.
However, they can be easily made equal to zero on the
axis of solenoid, where the beam is located. It is possible, at least 
in principle,
that only $y$-component of magnetic field appears, which does not produce
any effect.}. Integrating by part we obtain a following result:
\be
\dbv = -\frac{1}{Mc}\vE\times\hat{\vB}\int_{B_1}^{B_2}\pal (B)dB
\ee
where $\hat{\vB}$ is the unit vector along $\vB$ direction.
Notice that only the transverse to magnetic field polarizability
$\pal (B)$ enters expressions (6), (7) for the Hall drift velocity $\dbv$.
A rough estimate of $\dv$ magnitude can be found by substituting in (6)
$\pal \approx a^3$, where $a$ is the radius of the external atomic orbital.
In this way we arrive at an approximate formula:
\be
\dv \approx \frac{EBa^3}{Mc}
\ee
For $B=10T,\,\, E=3\cdot 10^3{V/cm},\,\, M=M_p$, ($M_p$ is the proton mass)
and $a=1\AA$ one finds from (8)
$\dv \approx 1.85\cdot10^{-5}cm/sec$. This is far beyond the experimental
resolution with no hope to increase the effect by changing fields. Even
more polarizable molecules have no chance to enhance the effect to a
remarkable value. However, the situation changes drastically for highly
excited  (Rydberg) atoms. This stems from the fact that $a\sim n^2$
where $n$ is the principal quantum number.
Therefore, the Hall velocity (8) is proportional to $n^6$.
For a modest value of $n=50$ the enhancement factor is $\sim 1.6\cdot 10^{10}$
producing $\dv\sim 2.89\cdot 10^{5}cm/sec.$
However, as we show below, a more accurate treatment reduces this value 
in approximately $n/m$ times where $m$ is the magnetic quantum number,
which is not large at a conventional way of excitation.
\par
An unexpected feature of the neutral atom Hall-effect is
that is does not depend on atomic path length in crossed fields. Instead
it depends on the variation of the electric or magnetic field or
polarizability.
Let us discuss eqn (7) in more details. We restrict our calculations
with a range of fields $B \ll c/n^3$ in atomic units.
This inequality guarantees that the diamagnetic energy $c^{-2}B^2n^4$
is much less than the Coulomb energy $n^{-2}$. For this range of fields
the transverse polarizability $\pal$ has been found
in \cite{BKKP}  to be:
\be
\pal(B)=\frac{9}{2}\frac{mn^{2}c}{B}
\ee
Here $m$ is the magnetic quantum number, which is well-defined in
external magnetic field. Plugging (9) into (7) we arrive at
\be
\dbv = -\frac{9mn^2}{2M}\ln\frac{B_2}{B_1}\cdot \vE\times\hat{\vB}
\ee
Two or three lasers can excite an atom from the ground-state to a
state with $m \leq 3$. Thus, in comparison to our apriori estimate
(8) the result given by eqn. (10) differs by a factor $(9/2)m(c/Bn^{4})$.
For the most interesting range of parameters the factor $c/Bn^{4}$ is
small (diamagnetic energy is much larger than the interlevel spacing).
Nevertheless, an estimate for the same
value of $E=3\cdot 10^{3}V/cm$, $M=M_p$ (Hydrogen),
$B_{2}/B_{1} \approx 2.7$, $m=3$ and $n=50$ eqn(10) gives $\mid \dv\mid\approx
2150 cm/sec$. For the distance between the magnet and the screen
$1 m$ and the initial velocity of
atoms $1 km/s$ the excited atoms form a spot at the distance $2.15 cm$ from the
central spot, produced by non-excited atoms. These figures  give a 
realization about expected magnititude of the Hall velocity. The
effect can be used to separate excited atoms from nonexcited ones and for
the measurement of $\pal(B)$. For $m=0$ the effect is much weaker. It is
determined by the diamagnetic Hamiltonian. The polarizability for
this case can be found in \cite{BKKP}.
\par
Let us analyze the contribution of the first term in the integral (4).
All fields depend on the coordinate $x$ only. On the other hand vector
$\vd$ is directed along $y$. Therefore, formally this term is equal
to zero. However, small angular misalignment $\delta\theta$ can
create a non-zero random contribution $\dv_{rand}$, proportional to
$\sim\frac{dE}{M\dv_0}(\delta\theta)$. For $E=3\cdot 10^{3}V/cm$, $B=10T$
the ratio $\dv_{rand}/\dv$ is $10^{-2}\delta\theta$ by the order of
magnitude and can be neglected.\\
{\it Alternating Fields}.
Consider an excited atom in an alternating electromagnetic field in a
resonator. We assume that the
alternating magnetic field $B_{x}=E\sin{\omega t}\sin{kz}$ is
directed along $x$-axis, the alternating electric field
$E_{y}=E\cos{\omega t}\cos{kz}$ is directed along $y$. The atomic
beam propagates along $x$-axis. The frequency
$\omega$ is assumed to be small enough to justify the adiabatic
approximation. To find the effect, we employ eqn (4) neglecting
the first term in the integral. In contrast to the case of permanent
fields, the main contribution to the Hall velocity in alternating fields
comes from the linear Stark effect, specific for the Hydrogen and Rydberg states.
Indeed, in this case the dipolar moment $\bf d$ in the electric field $\bf E$
has a constant modulus, but changes its direction with $\bf E$:
\be
{\bf d}=d_{nk}\hat{\vE}
\label{LStark}
\ee
where $d_{nk}=(3/2)nk$ in atomic units is the dipolar moment due to the linear
Stark-effect, $k$ is the parabolic quantum number associated with the
quantized Runge-Lenz vector $\bf A$
and varying from $-(n-\mid m\mid -1)$ till $n -\mid m\mid -1$
\cite{LL}; $\hat{\vE}$ is the unit vector parallel to $\bf E$. In our geometry
the vector $\hat{\vE}$ is always collinear to the $x$-axis and changes its sign
each half-period. Thus:
\be
\dot{\vd}\,=\,3nk\hat{x}\sum_{n=-\infty}^{\infty}(-1)^n\delta\left(
t-{(n-1/2)\pi\over\omega}\right)
\label{ddot}
\ee
Integrating $\dot{\vd}\times\vB$, according to Eqs.(4) and (\ref{ddot}),
and neglecting oscillating terms, we find:
\be
\dv_z\,=\,{3\omega nkEl\over \pi Mcv_0}\sin kz\, {\rm sign}(\cos kz)
\label{dv-linear}
\ee
where $l$ is the length of the path passed by atoms in the $ac$ fields
and $v_0$ is the velocity of atoms in the beam.
We have assumed that the atomic beam width is much smaller than the 
wave-length of
the $EM$-wave. The maximal effect is reached at $\sin kz =1$.
For $\omega = 2\pi 10^{8} sec^{-1}$, $E = 100 V/cm$, $l=10 cm$,
$v_0 =1 km/sec$, $n=k=50$ we find $\dv_z = 6,700 cm/sec$.
Note that the $ac$ Hall-effect is
proportional to the path length $l$, passed by the atoms in the $ac$ field.
This fact provides an additional opportunity to enhance the
effect.
\par
An important question is: what values of the quantum number $k$ can be
reached at the excitation of Rydberg states. We calculated matrix elements
for dipolar transitions from the ground state to a highly excited state
with the parabolic quantum numbers $n_1$ and $n_2$ and have found that
they decrease slowly with $n_i$ ($\sim n_i^{-3/2}$). Therefore the
exitation of large $k=n_1-n_2$ is reasonably probable.
\par
In our calculations we did not take into account the finite size of the atomic
remainder, which lifts the degeneracy over the total orbital moment $l$.
Thus, if the
frequency is large in comparison to the finite size splitting
$\delta E\sim n^{-6}$ for $s$-wave the linear Stark effect is
responsible for the effect. Otherwise the quadratic Stark-effect
should be accounted for. For $n=50$ the crossover frequency is about
$1 MHz$. The quadratic effect is much weaker. Details will be published
elsewhere.

{\it Limitations of the effect}.
There are two kinds of limitations for the Hall-effect magnitude:
intrinsic limitations, wich stem from physical reasons, and methodical
limitations associated with the range of validity for the theory i.e. with
what we can calculate.
An example of an intrinsic limitation is given by inequality 
$\dv <4.5n^3E\ln (B_2/B_1)$, following from eqn. (10).
Another limitation $\omega <\frac{1}{n^3}$ means that at higher frequencies
atomic electrons easily transit from one orbit to another and the adiabaticity
is violated. It seems to be a kind of methodic limitations, since our description
definitely fails at so high frequencies. On the other hand, for such a large
frequency the atomic electrons on remote orbits do not have time enough
for polarization. Thus, we expect the effect to decrease rapidly at
$\omega >1/n^3$.
\par
An important methodic limitation is $B\ll\frac{c}{n^3}$.
Expressions for polarizability etc., which we have employed earlier, are
valid in this range of $B$. However, it does not mean that the effect
vanishes at higher fields. It does exist, though we can not describe it
properly. The range $B\sim c/n^3$ is of a great interest 
\cite{delande},\cite{Kleppner}, since it
is a range of the quantum chaos, where the levels are located randomly according
to the orthogonal ensemble \cite{ensembles}. The problem of
polarizability of random quantum levels is still open. It is worthwile to 
note
that the experimental data on atomic drift in this range of fields could give
an important information on the properties of quantum states.
\par
An important question is the validity of the adiabatic approximation.
For static fields a necessary requirement is that the field variation
in a frame of reference, associated with moving atoms, is slow in
the scale of atomic frequency $1/n^3$ or even the smaller diamagnetic
frequency  $B^{2}n^{3}/c^2$. The size of the edge inhomogeneity for 
electric field is
determined by the distance $h$ between condenser plates. Accepting $h=0.5cm$
and $\dv_0 = 10^{5}cm/s$ we see that even the second requirement is satisfied
reliably: $h/\dv_0\ll B^2n^3/c^2$.
\par
The adiabatic approximation fails when an atom transits from the ground state to
an excited state. However, the final state is established in the interval of time
equal to the inverse transition frequency for this state by the order of 
magnitude.
Therefore, one can neglect
this interval, very short in the scale of the center-of-mass motion. Due to
a long life-time of the Rydberg atoms (about $0.1 sec$ for $n=50$), one can
neglect the possibility of reciprocal transitions to the ground state on the way
of a Rydberg atom from the condenser to the observation point.
\par
A serious obstacle for observation of $ac$ fields effect may
the fact that a strong $ac$ field produces ionization of Rydberg atoms rather
effectively \cite{ionization}. Therefore, the length of the resonator $l$ should
be small enough
to reduce the ionization probability. On the other hand, the drift velocity is
proportional to $l$. Hopefully, ajusting the field parameters $n$, $E_1$, and
$l$ one can find a range where the $ac$ Hall-effect is observable.
However, an additional analysis is necessary.

In conclusion we propose to find experimentally a new phenomenon:
Hall-effect (drift) of neutral atoms. The best objects for observation of this
effect are highly excited (Rydberg) atoms, due to their gigantic polarizability.
Another possibility may be eximeric molecules. For experiments with
static fields simple estimates made in the text show that effect is not small.
In a most natural geometry Hall velocity is perpendicular to the initial
velocity of atoms. This deviation enables one to separate excited atoms.
Measurements of the Hall velocity would be a source of additional information on the
spectrum and wave functions of atomic electrons especially in the quantum chaos 
range $B\sim c/n^3$. In the case of static fields the theory predicts
that the drift velocity does not depend on the time which atoms spend in
fields, but only on spatial variations of fields.
\par
The atomic drift can be observed also in alternating fields forming standing
waves. In this situation it grows linearly with the path-length passed in the
$ac$ field and its amplitude.

One of the authors (V.L.P.) is grateful to G. Welch for an illuminating 
discussion on the experimental situation and indicating several references.

\end{document}